\documentclass[11pt]{article}
\usepackage{moriond,epsfig}

\def\be{\begin{equation}}
\def\ee{\end{equation}}
\def\bea{\begin{eqnarray}}
\def\eea{\end{eqnarray}}

\begin{document}
\vspace*{4cm}
\title{Studies of Top Quark properties at the LHC}

\author{ S. Bentvelsen }

\address{NIKHEF \\ Kruislaan 409 \\ 1098 SJ Amsterdam \\ the Netherlands}

\maketitle

\abstracts{ The LHC collider plans to start in 2007, after which
millions of top quarks will be produced at the collision points.  In
the first period of data taking these events will provide an essential
calibration tool for the ATLAS and CMS detectors. During a later stage, as
part of the precision measurement programme, both detectors will
determine the top mass and properties like spin, charge and couplings.
Top physics also plays a central role in the search for new particles,
both as source of a possible signal and as source of background.}

\section{Top quark physics}
There are many motivations to study top quarks at LHC.  It is by far
the heaviest fundamental particle, with a mass close to the
Electro-Weak Breaking Scale (EWBS). Precise determination of its mass allows
stringent tests on the Standard Model (SM) and constrains via radiative
corrections the mass of the Higgs boson~\cite{Beneke:2000hk}. In many scenarios
beyond the SM heavier particles decay into top quarks.
Studying detailed properties of top quarks can give handles on new
physics.

However also from the experimental point of view top quarks are
important. Since they are abundantly produced at LHC, they provide an
essential tool for understanding the detectors during the commissioning
phase. Triggering, tracking, $b$-tagging, energy and jet calibration 
can all benefit from the present top signal. 

Lastly, top quarks will be a major source of background for almost
all searches for physics beyond the SM. Precise
understanding of the top signal is crucial to claim new physics.

The cross section for $t\bar{t}$ production has been calculated up to
NLO order, and results in $830 \pm 100$ pb$^{-1}$, where the
uncertainty reflects the theoretical error obtained from varying the
renormalisation scale by a factor of two. With a luminosity of 10
fb$^{-1}$ in the first years of LHC running, and reaching 100 fb$^{-1}$
at the higher luminosity stage, top physics will hardly be limited by
statistics.

The partonic center of mass (CM) energy $\sqrt{\hat{s}}$ has to exceed
twice the top mass $m_t$, and therefore the production dominates at $\hat{s} =
s x_1 x_2 \sim (350 \mbox{ GeV})^2$, which corresponds to $x_1 x_2 \sim
10^{-3}$ for the LHC CM energy of 14 TeV. 
The gluon density of the proton dominates at these values of
$x$ and about 87\% of the $t\bar{t}$ pairs is expected to be produced
by gluon fusion, while the remaining 13\% via quark-anti-quark
annihilation (at the Tevatron, with a CM energy close to 2 TeV, the
$t\bar{t}$ cross section is approximately a factor of 100 smaller, and
the bulk of $t\bar{t}$ pairs are produced via the quark-anti-quark
annihilation process).

\section{Top Mass measurements}
The inclusive lepton plus jet channel, $ t \bar t \rightarrow WWb \bar b 
\rightarrow (l \nu) (jj) (b \bar b) $, provides a large and clean
sample of top quarks and is
the most promising channel for an accurate determination of
the top quark
mass~\cite{Borjanovic:2004ce,Sonnenschein:2000wg}. Considering only
electrons and muons, the branching ratio of this channel is 29.6\%. 
Various methods
have been
exploited to measure the top quark mass. In the most straightforward
method the hadronic part of the decay is used, and the top mass is
obtained from the invariant mass of the three jets coming from the
same top: $m_t=m_{jjb}$.

The typical selection of single lepton top events is based on the
presence of an isolated high $p_T$ lepton with $p_T>20$ GeV and 
missing energy $E_T^{miss}>20$ GeV.
At least four jets, typically reconstructed with a cone size of $\Delta R =0.4$,
with $p_T>40$ GeV and $|\eta|<2.5$ are required. One or two jets are required to be tagged as
$b$-jets. The reconstruction of the decay $W\rightarrow jj$ is first
performed by selecting the pair of non $b$-tagged jets with invariant mass
closest to $m_W$. Events are retained only if $|m_{jj}-m_W|<20$
GeV. The combination of the jet pair $jj$ with the $b$-tagged jet
yields a combinatoric ambiguity. For events with only one tagged $b$-jet the
events are kept for which the opening angle of the $b$-jet with the $W$
is smaller than with the lepton of the event. For events with two
$b$-tagged jets, the $b$-jet which resulted in the highest $p_T$ of
the system was combined with the jet pair $jj$.   

\begin{figure}[th]\label{f:semilep}
\begin{center}
\psfig{figure=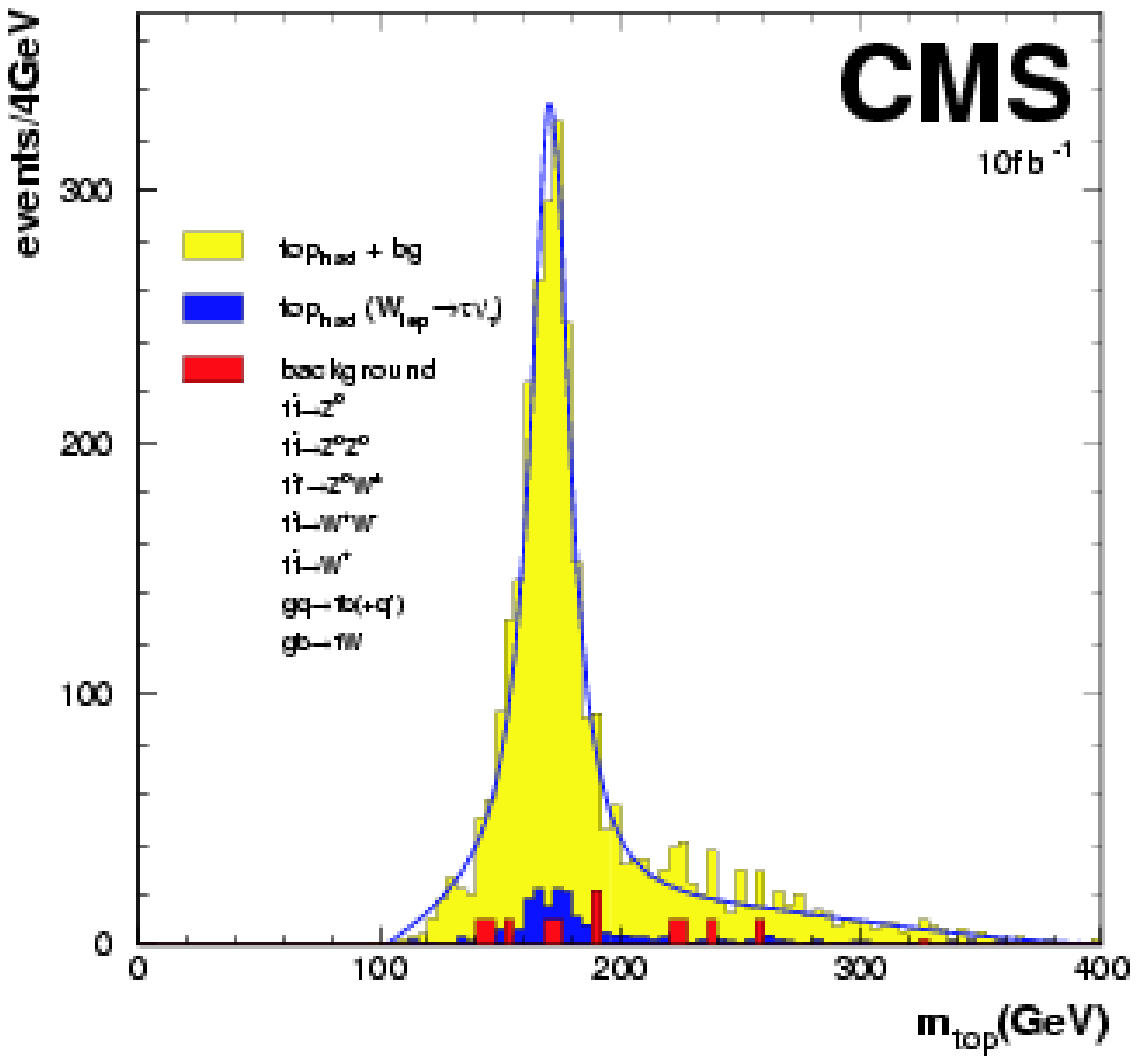,height=45mm}
\psfig{figure=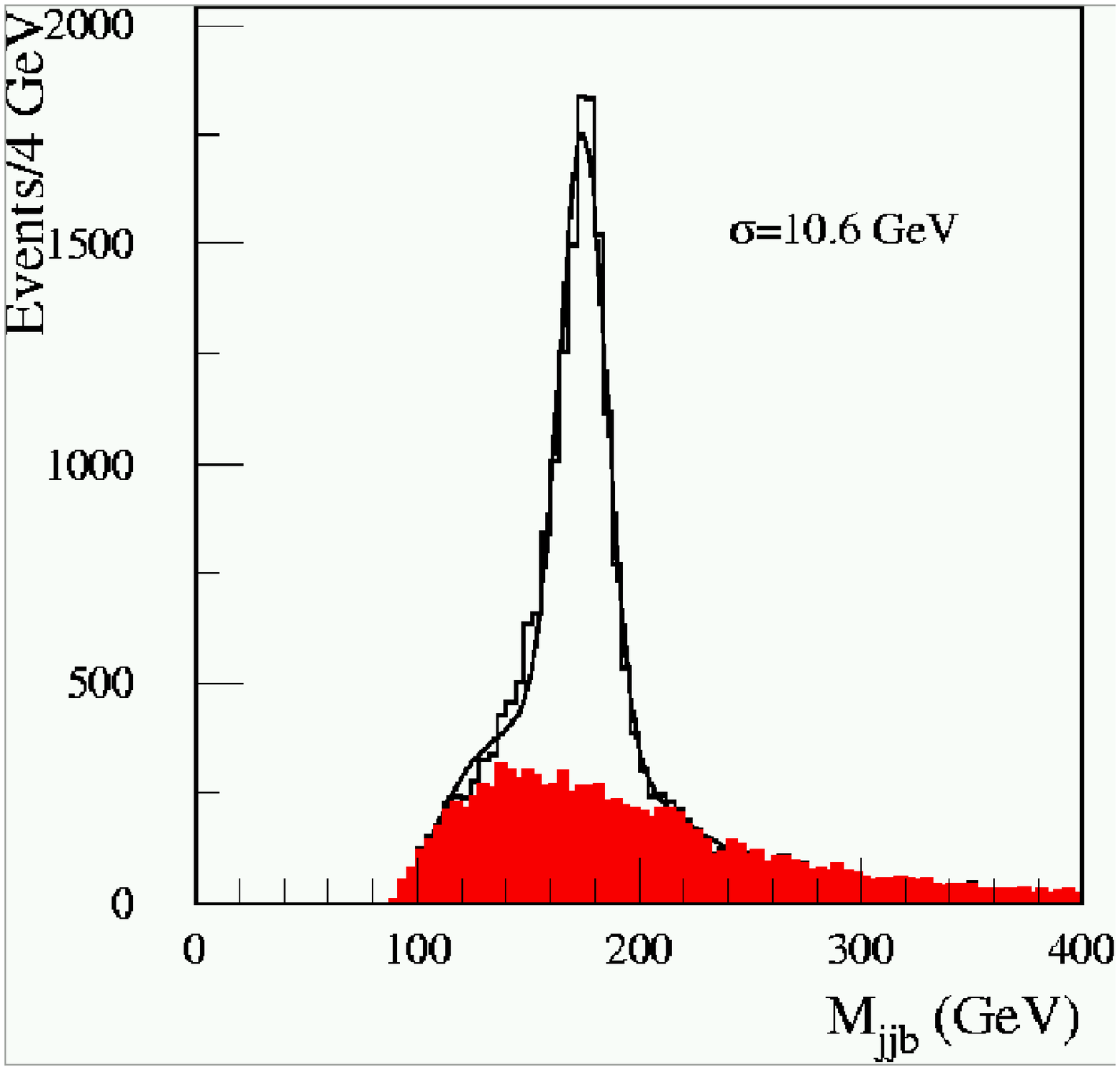,height=45mm}
\caption[]{Reconstructed top mass for the CMS (left) and ATLAS (right)
  collaborations, for 10 fb$^{-1}$ of data.}
\end{center}
\end{figure}
In Figure~\ref{f:semilep} the reconstructed $m_t$ for the ATLAS and
CMS detectors is shown. The statistical uncertainty on the top mass is
not a problem and the background is 
well under control, with a signal to background ratio $S/B\sim 65$.

 The largest systematic uncertainties arise from
the jet energy scale, the $b$-quark fragmentation, the initial and
final state radiation and the background contributions. The studies
indicate that a total error on $m_t$ below 2 GeV should be feasible,
possibly reaching an ultimate precision around 1 GeV.

\begin{figure}[th]\label{f:comtop}
\begin{center}
\psfig{figure=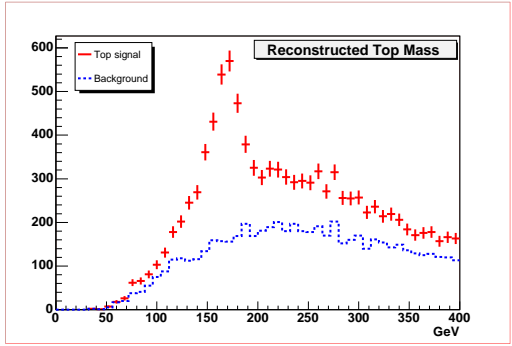,height=45mm}
\caption[]{Reconstructed top mass, without $b$-jet tagging, for 150
  pb$^{-1}$ of data. The background corresponds to $W$+4 jet events. 
} 
\end{center}
\end{figure}
Already in the commissioning phase of the detectors, 
during the startup of LHC, it will be possible to observe a top
signal. ATLAS performed a study in which $b$-jet
tagging is assumed to be absent, as pessimistic scenario. In this case
the top 
mass is reconstructed from a sample of events with exactly four
reconstructed jets and the three jets which result in the
highest invariant $p_T$ are used to calculate the invariant mass. 
Figure~\ref{f:comtop} shows the expected signal for a luminosity of
150 pb$^{-1}$, i.e. after a few days of running. The most important
background, production of $W + 4 $ jets, is determined from matrix
element calculations using the AlpGen event generator\cite{alpgen},
whereas the signal is obtained from the NLO Monte Carlo generator
MC@NLO\cite{mcatnlo}.  


\section{Top properties}
The high statistics available at LHC, will allow to study many
properties of the top quark.
In order to confirm that its electric charge is
indeed $Q_{top}=2/3$, one can either measure the charge of the
$b$-jet and the $W$ boson, or attempt to directly measure the top
quark coupling through photon radiation in $pp\rightarrow
t\bar{t}\gamma$ and $pp\rightarrow t\bar{t}$ with $t\rightarrow
Wb\gamma$. Since the first process is dominated by $gg$ fusion at LHC,
one expects that the $tt\gamma$ cross section is approximately
proportional to $Q^2_{top}$.

The 
treatment of the radiative top
production and top decay
matrix elements, fed into the Pythia Monte Carlo, was based on the
on-mass approach for the decaying top, i.e. the production and decay
were treated independently. By suitable selection criteria the hard
$\gamma$ radiation from top production can be enhanced. It 
is foreseen to disentangle the top charge
during the first year of running at LHC using this method~\cite{topcharge}.

The $t\bar{t}$ spin correlations allow the direct measurement of the
top quark spin. In the dileptonic $t\bar{t}$ channel the helicity
angles $\theta^*$ of the leptons are given by the double differential
cross section
\be
\frac{1}{\sigma_{tot}} \frac{d^2 \sigma}{d \cos \theta_{l^+}^* d \cos
  \theta_{l^-}^*} = \frac{1}{4} \left( 1-{\cal A} \cos \theta_{l^+}^*
  \cos \theta_{l^-}^* \right)
\ee
with the asymmetry ${\cal A}$ in the helicity basis defined as the
normalized difference between like-spin and unlike-spin top pairs. At
the LHC where $gg$-fusion dominates, the asymmetry is expected to be
${\cal A}=0.31 \pm 0.03$\cite{topspin}. Studies are underway to
determine also in semi-leptonic top events the
top spin correlations. In these studies the 
softest jet is used as spin analyzer.

\section{Exotic models}
\label{section:exo}
\begin{figure}[th]
\begin{center}
\psfig{figure=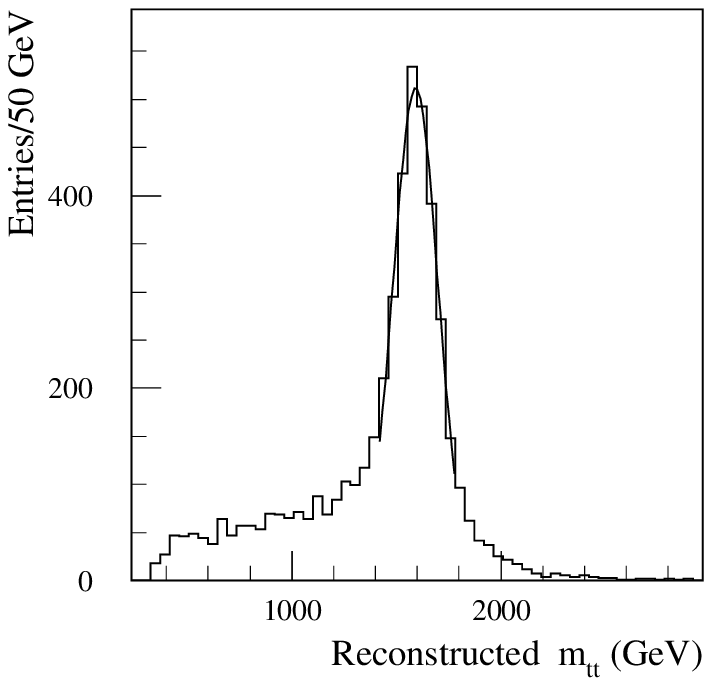,height=40mm}
\psfig{figure=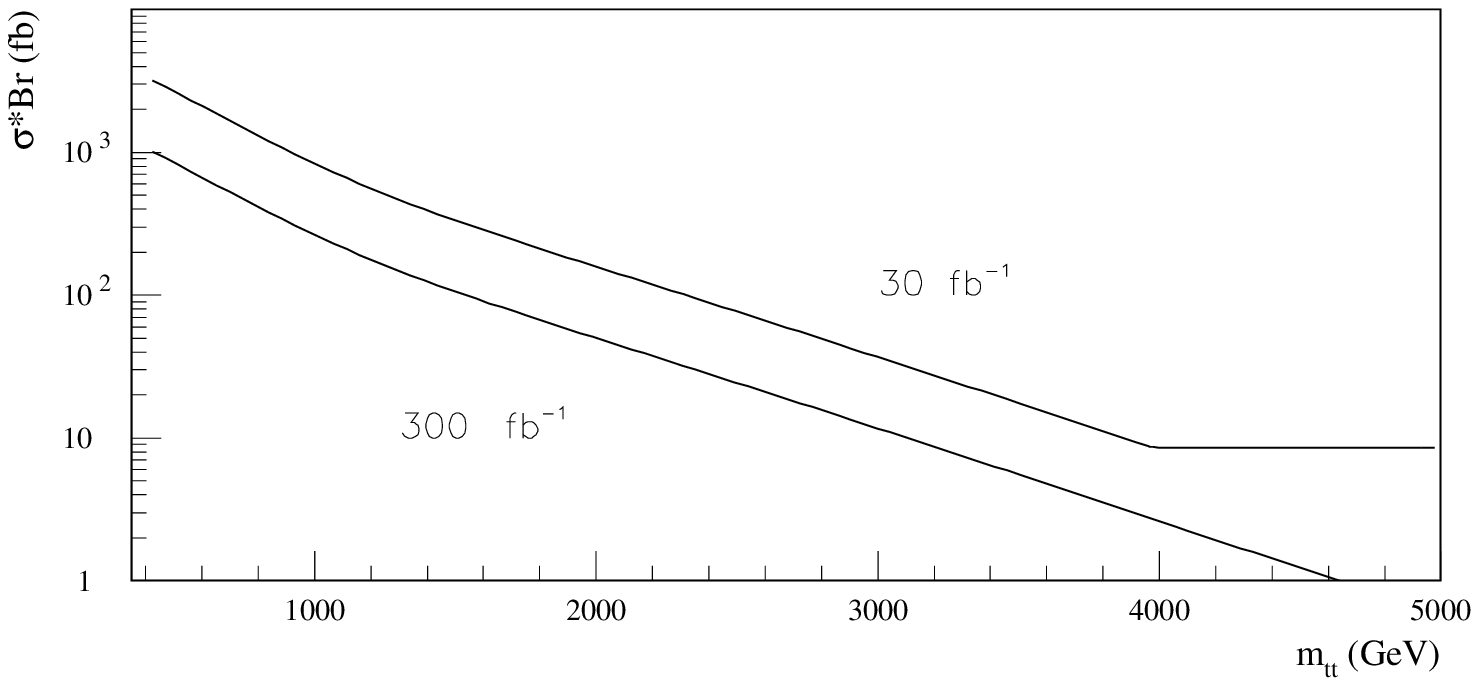,height=40mm}
\caption{On the left: Measured $t \bar t$ invariant mass distribution for
reconstruction of a narrow resonance of mass 1600 GeV decaying to $t \bar
t$. On the right: Value of $\sigma$xBR required for a 5$\sigma$ discovery
potential
 for a narrow resonance decaying to $t \bar t$, as a function of m$_{t
\bar t}$.}
 \label{peak} 
\end{center}
\end{figure}

Due to its large mass, the top quark could be part of massive 
particle decays. Its clear experimental signature 
makes it a very interesting tool to study
 the exotic decays. Some examples include: "Heavy top" in Little Higgs
models, signatures which include the quark top in models with
extra-dimensions, search of resonances which decay in $t \bar t$ (as
predicted in SM Higgs, MSSM Higgs, Technicolor models, strong electroweak
symmetry breaking models, Topcolor, etc.).
 Physics beyond the SM could affect cross section measurements for 
$t \bar t$ production in a variety of ways: a
heavy resonance decaying to $t \bar t$ might enhance the cross section, 
 and might produce a peak in the $t \bar t$ invariant mass spectrum.
Because of the large variety of models and their parameters, in ATLAS a
study was made~\cite{Beneke:2000hk} of the sensitivity to a "generic" narrow
resonance decaying
to $t \bar t$. Events of the single lepton topology
were
selected. In addition, between four and ten jets were required with p$_T$$>$20 GeV
and $|$$\eta$$|$$<$3.2, with at least one of them
tagged as $b$-jet. After these cuts, the
background is dominated by the $t \bar t$
continuum.
The obtained mass resolution $\sigma$ (m$_{t \bar t}$)/m$_{t \bar t}$ was 
equal to
6.6$\%$. 
As an example, Figure~\ref{peak} shows the reconstructed m$_{t \bar t}$
distribution for a narrow resonance of mass 1600 GeV.

The reconstruction efficiency, not including BRs, was about 20$\%$ for a
resonance of mass 400 GeV, decreasing gradually to about 15$\%$ for m$_{t
\bar t}$ = 2 TeV.
For a narrow resonance $X$, fig.~\ref{peak} shows the required
$\sigma$xBR($X$$\rightarrow$$t \bar t$) for a discovery (the signal 
must have a statistical significance of at least
5$\sigma$ and must contain at least 10 events). Results are shown as a
function of m$_X$ for integrated luminosity of 30 fb$^{-1}$ and 300
fb$^{-1}$.

\section{Rare decays}
\label{section:rare}
With its large mass, the top quark will couple strongly to the EWSB
sector. Many models of physics beyond the SM include a more
complicated EWSB sector, with implications for top quark
decays. Example includes the possible existence of charged Higgs
bosons, or possibly large flavour changing neutral currents (FCNC)
in top decays. In the SM, FCNC decays of the top quark are highly
suppressed (BR $<$ 10$^{-13}$-10$^{-10}$). However, several extensions
of the SM can lead to very significant enhancements of these BRs
(10$^{-3}$-10$^{-2}$ or even higher). The sensitivity to some of these
scenario~\cite{Beneke:2000hk} has been investigated by both ATLAS and CMS.

In particular the FCNC processes $t \rightarrow Zq $, $t \rightarrow
\gamma q$, $t \rightarrow gq$, $t \rightarrow WbZ$, $t \rightarrow
WbH$, $t \rightarrow Hq$ have been studied. It has been shown that the
limit on the branching ratios of these processes can be improved by
orders of magnitude with respect to the current limits, and will range from
$10^{-7}$ for $t \rightarrow WbZ$ to $10^{-3}$ for $t \rightarrow gq$
using 100 fb$^{-1}$ of data.

\section{Single top production}
\label{section:stp}
\begin{figure}[h]
\centerline{
\includegraphics[width=12cm]{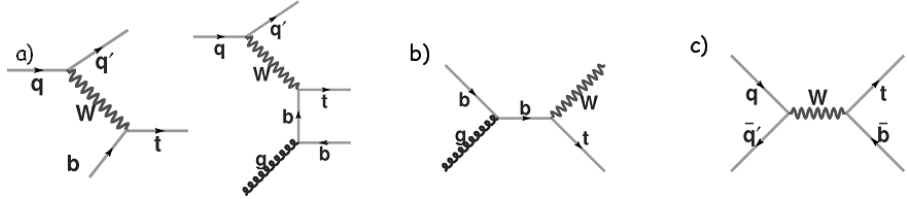}}
\caption{Feynman diagrams for the electroweak single top quark processes
accessible at the LHC: a) $W$-gluon fusion, b) $Wt$ production, c)
s-channel or $W^*$ process.}
\label{single} 
\end{figure}

The precise determination of the properties of the $W$-$t$-$b$ vertex, and
the
associated coupling strengths, will more likely be obtained from
measurements of the electroweak production of single top quarks.
Single top quarks can be produced via three different reactions
 (shown in Figure~\ref{single}): $W$-gluon
fusion ($\sigma$ $\simeq$ 250 pb), $Wt$ production ($\sigma$ $\simeq$
 60-110 pb) and $W^*$ process ($\sigma$ $\simeq$ 10 pb). 
There are important backgrounds with final states similar to the signals
under study ({\it e.g.} $\sigma(t \bar t)$ =830 pb, $\sigma(Wb \bar b)$
$>$300 pb)
 and the possibility to extract a signal depends critically from 
 the detector performance. Important parameters are the rate of fake
leptons,
the $b$-jet identification and fake $b$-jet rate, the capability to
identify
forward jets and to veto low energy jets.
To reduce the enormous QCD multi-jets 
background, the $t \rightarrow l \nu b$ decay has been studied, and a 
 pre-selection was done in which a high p$_T$ lepton, at least two jets,
at least one $b$-tagged jet, and one forward jet are required.
 It is interesting to study the three processes
separately, since
they have separate sets of backgrounds, their
systematic errors for V$_{tb}$ are different, and they are differently
sensitive to new physics.
For example, the presence of a heavy $W$  would result in an increase of
the $W^*$ signal. Instead, the existence of a $FCNC$ $gu \rightarrow t$
would
be seen in the $W$-gluon fusion channel.
Discriminants for the three signals are for example: the jet multiplicity
(higher for $Wt$), the presence of more than one jet tagged as a b (this
increase the $W^*$ signal with respect to the $W$-gluon fusion one), the
mass distribution of the 2-jet system (which has a peak near
the $W$ mass for the $Wt$ signal and not for the others).

The results found for the relative experimental statistical errors
on the production cross sections of the single top processes, 
 imply statistical uncertainties on the extraction of V$_{tb}$ of
0.4$\%$ for $W$-gluon fusion, 1.4$\%$ for $Wt$ and 2.7$\%$ for $W^*$.
 The errors in the extraction of  V$_{tb}$ would be dominated by
uncertainties in the theoretical predictions of the cross-sections. These
arise from uncertainties in the parton distribution structure functions
(PDFs), uncertainty in the scale ($\mu$)  used in the calculation,
 and the experimental error on the top mass, and amounts to
 approximately 6$\%$ for all three processes.

\section{Conclusions}
At LHC, the top mass will be determined with an ultimate
precision of approximately 1 GeV. However, already during the commissioning
phase of the LHC the top quark signal will be observed using
simple and robust reconstruction methods. 

The high statistics sample will allow to determine top quark properties
like charge and spin unambiguously. Exotic models involving top will
be tested with high precision. Limits on rare FCNC top decays
will be improved by orders of magnitude, or even observed experimentally. 
The observation of single top production will allow a precise determination 
of $V_{tb}$.


\section*{References}
%
%

\bibliography{/group/atlas/users/stanb/private/writeup/Bib/spires}

\begin{thebibliography}{1}

\bibitem{Beneke:2000hk}
M.~Beneke et~al.,
\newblock Top quarks physics,
\newblock {\em hep-ph/0003033}, 2000.

\bibitem{Borjanovic:2004ce}
I.~Borjanovic et~al,
\newblock Investigation of top mass measurements with the ATLAS detector at
  LHC,
\newblock {\em hep-ex/0403021}, 2004.

\bibitem{Sonnenschein:2000wg}
L.~Sonnenschein,
\newblock Top quark physics at the lhc.
\newblock {\em Prepared for 14th International Spin Physics Symposium (SPIN
  2000), Osaka, Japan, 16-21 Oct 2000}, 2000.

\bibitem{alpgen}
M.L. Mangano, M. Moretti, F. Piccinini, R. Pittau, A. Polosa,
\newblock Alpgen, a generator for hard multiparton processes in hadronic
  collissions,
\newblock {\em JHEP 0307:001}, 2003.

\bibitem{mcatnlo}
S.~Frixione,~B. Webber,
\newblock Matching NLO QCD computations and parton shower simulations,
\newblock {\em JHEP06}, 2002.

\bibitem{topcharge}
M.~Ciljak et~al.,
\newblock Top charge measurement at the ATLAS detector,
\newblock {\em ATL-PHYS-2003-035}, 2003.

\bibitem{topspin}
K.~Smolek and V.~Simak,
\newblock Measurement of spin correlations of top-antitop pairs,
\newblock {\em ATL-PHYS-2003-012}, 2003.

\end{thebibliography}
%


\end{document}